\documentstyle[12pt,amssymbols]{article}

\def \be {\begin{equation}}
\def \eq {\end{equation}}
\def \bee {\begin{eqnarray}}
\def \eqq {\end{eqnarray}}

\def \bea {\begin{array}{c}}
\def \eqa {\end{array}}

\def \ra {\rangle}

\def \dels {\partial\kern-.5em / \kern.5em}
\def \As {{A\kern-.5em / \kern.5em}}
\def \Ds {D\kern-.7em / \kern.5em}
\def \Psib {\bar{\Psi}}

\def \H {{\cal H}}

\def \a {\alpha}

\def \G {\Gamma}

\def \m {\mu}
\def \n {\nu}
\def \k {\kappa}

\def \th {\theta}
\def \Th {\Theta}

\def \dag {\dagger}

\def \Psib {\bar{\Psi}}

\def \II {I\hspace{-.1em}I\hspace{.1em}}
\def \IIA {\mbox{\II A\hspace{.2em}}}
\def \IIB {\mbox{\II B\hspace{.2em}}}
\def \yd {y^{\dagger}}
\def \thd {\th^{\dagger}}

\def \chid {\chi^{\dagger}}

\def \one {{\bf 1}}

\begin{document}
\begin{titlepage}
\today          \hfill
\begin{center}
\hfill    UU-HEP/97-03\\

\vskip .5in

{\Large\bf Brane Creation in M(atrix) Theory}

\vskip .5in
Pei-Ming Ho\footnote{Address after September 1, 1997:
Department of Physics, Jadwin Hall, Princeton University,
Princeton, New Jersey 08544, U.S.A.}
and Yong-Shi Wu\footnote{Address after September 1, 1997:
School of Natural Sciences, Institute for Advanced Study,
Olden Lane, Princeton, New Jersey 08540, U.S.A.}\\
\vskip .3in
{\em Department of Physics,
University of Utah \\
Salt Lake City, Utah 84112, U.S.A.}

\end{center}

\vskip .5in

\begin{abstract}

We discuss, in the context of M(atrix) theory,
the creation of a membrane suspendend between two longitudinal
five-branes when they cross each other.
It is shown that the membrane creation is closely related to
the degrees of freedom in the off-diagonal
blocks which are related via dualities to the chiral
fermionic zero mode on a 0-8 string.
In the dual system of a D0-brane and a D8-brane in type \IIA theory
the half-integral charges associated with
the ``half''-strings are found to be connected to the
well-known fermion-number fractionalization
in the presence of a fermionic zero mode.
At sufficiently short distances,
the effective potential between the two five-branes
is dominated by the zero mode contribution
to the vacuum energy.

\end{abstract}
\end{titlepage}

\newpage
\renewcommand{\thepage}{\arabic{page}}
\setcounter{page}{1}
\setcounter{figure}{0}

%THIS IS PAGE 1 (INSERT TEXT OF REPORT HERE)

\section{Introduction} \label{intro}

It was first suggested by Hanany and Witten \cite{HW}
that when two branes (in appropriate configuration)
cross each other, a third brane stretching between them
is created or annihilated.
They observed the creation in type \II B string
theory of a D3-brane when an NS5-brane
crosses a D5-brane. It is related to
creation of branes of other dimensions by
sequences of dualities \cite{HW,BDG,DFK,BGL,dAl}.
In particular it is dual to the creation of a
fundamental string when a D0-brane crosses
a D8-brane in type \II A theory, as well
as the creation of a membrane in M theory
when two five-branes sharing one common
dimension cross each other. In \cite{BDG}
it was argued from the anomaly equation
that when the two branes in question cross
each other an energy level crosses zero,
and thus a single particle or hole is created.
The creation of a particle is understood as
the creation of an open string or brane.
In \cite{DFK} it was also shown that
the induced charge
on the D8-brane world-volume indicate the creation
of a string when the D0-brane crosses the D8-brane.

In this paper we try to understand the
phenomenon of brane creation in the
context of M(atrix) theory \cite{BFSS}.
It is most convenient to consider the creation
of a longitudinal membrane when crossing
two longitudinal five-branes. The classical
background of the longitudinal five-branes
are described by topologically nontrivial
gauge field configurations residing in
diagonal blocks \cite{BSS}. We will show
that the membrane creation is closely related to
the degrees of freedom in the
off-diagonal blocks.
In accordance with T-duality, the equations
of motion for the off-diagonal blocks in
the background of two longitudinal
five-branes are formally the same as
the case of the 0-8 string.
In a previous paper \cite{HLW},
the existence of the chiral
fermionic zero mode on the 0-8 string
has been derived, using the index theorem,
from the classical equations of motion
in M(atrix) theory for the off-diagonal
blocks in the background of diagonal ones.
It is this
chiral fermionic zero mode in the
off-diagonal blocks that gives rise to
the energy level, which crosses zero,
proposed in \cite{BDG} (Sec.\ref{zero}).
The energy of this fermionic mode is
linear in the distance between the five-branes
and the proportional factor equals the membrane
tension so that upon quantization
it can be associated with the
production of membrane (Sec.\ref{SQ}).
The fact that the induced charge on the D8-brane
world-volume is $\pm 1/2$, as a result of proper
operator ordering, is closely related
to the well-known fractionalization of fermion
number \cite{JR} due to the existence of a fermionic zero mode
or mid-gap mode (Sec.\ref{FermNum}).
We will also argue that
when the five-branes are sufficiently close to
each other, the effective potential between
them is dominated by the contribution from the
off-diagonal degrees of freedom associated with
this zero mode (Sec.\ref{potential}).
Our results agree with string theory calculations
\cite{Lifs,BGL} for the effective potential
between a D0-brane and a D8-brane in type \IIA theory.
Our study indicates that in the M(atrix) model
the description of creating an open
membrane stretching between
two five-branes necessarily involves second quantization
of degrees of freedom residing in off-diagonal blocks.

\section{Fermionic Zero Mode in Off-Diagonal Blocks}
\label{zero}

The M(atrix) theory is defined by the action \cite{BFSS}
\be
S=\int dt\; Tr\left(\frac{1}{2R}\dot{X}_i^2
+\frac{R}{4}(2\pi T_2^M)^2[X_i,X_j]^2
+\frac{i}{2}\Psi^{\dag}\dot{\Psi}
+\frac{R}{2}(2\pi T_2^M)\Psib\G^i[X_i,\Psi]\right),
\eq
where $i,j=1,2,\cdots,9$,
$R$ is the radius of the eleventh dimension
and $T_2^M$ is the membrane tension.

Consider in M theory two five-branes
lying in directions (1,2,3,4,11) and directions
(5,6,7,8,11), respectively. In M(atrix) theory,
the $11$-th direction is the longitudinal
direction for an infinite momentum frame, and
the above configuration is described by
matrices in the block form:
\be
X_{\m}=\left(\begin{array}{cc}
                Z_{\m} & y_{\m} \\
                \yd_{\m} & W_{\m}
             \end{array}\right), \quad
\Psi=\left(\begin{array}{cc}
                \Th & \th \\
                \th^{\dag} & \psi
             \end{array}\right). \quad
\eq
The five-branes are residing in the diagonal blocks
and treated as background: We take $Z_a=0$ for
$a=5,\cdots,8$, $W_i=0$ for $i=1,\cdots,4$,
$Z_9=0$ and $W_9=x_9\one$, while
$Z_i$ and $W_a$ are realized as covariant
derivatives with topologically nontrivial gauge field
configurations on two four-tori $T^4$'s \cite{BSS}.
(We will use $i,j,k,\cdots$ for
the values $1,2,3,4$; $a,b,c,\cdots$ for $5,6,7,8$
and $\m,\n,\k$ for $0,1,\cdots,9$.)
Superpartners of $Z_{\m}$ and $W_{\m}$
($\Th$ and $\psi$) are set to be zeros.
The coordinate $x_9$ gives the transverse
distance between five-branes.
The variables $y_{\m}$ and $\th$ in the
off-diagonal blocks, dependent on the coordinates
of the torus $T^8=T^4\times T^4$, represent
the matrix model degrees of freedom which are
analogues of open strings between D-branes
in string theory.
We will treat them quantum
mechanically, and their fluctuations will
give rise to the interactions between the
BPS branes.

By compactifying dimensions $(1,2,3,4,11)$,
one can use dualities to relate the creation of a membrane by
crossing two five-branes in M theory
to the creation of a string by crossing
a D0-brane with a D8-brane in \IIA theory.
Thus we can compare our results about the former
with string theory calculations for the latter.

The M(atrix) theory action induces an action
for $y$ and its superpartner $\th$ in the
background of $Z_i$ and $W_a$. It is easy to
see that in accordance with T-duality,
this action for $y$ and $\th$ is
formally the same as that for a 0-8 string
(in the strongly coupled type \IIA theory)
which we derived previously in \cite{HLW},
with the $U(K)$ covariant derivatives
\footnote{The integer $K$ is proportional to
the longitudinal momentum of the five-branes
\cite{HLW}, analogous to the length of the
conjugacy class for a long string in the
matrix string theory \cite{BS,DVV}.}
$D_i=i Z_i$, $D_a=-i W_a$ defined on a dual
eight-torus $T^8$, except that now the distance
between the D0-brane and D8-brane is $x_9$
instead of zero.
The action for $\th$ is
\be
L_F=\int\th^{\dag}\left(i\dot{\th}
-2\pi R T_2^M(i\sum_{\mu=1}^{8}\G^{0\mu}
D_{\mu}+\G^{09}x_9)\th\right),
\eq
which is integrated over the dual $T^8$.

An example for the five-brane
configuration is given by
$[D_{2n-1},D_{2n}]=-if$ with a constant
$f=2\pi R_{2n-1}R_{2n}/K$ for $n=1,2,3,4$
\cite{BSS,HLW}.
When $x_9=0$, it was shown for this example \cite{HLW}
that there is only one chiral fermionic zero
mode for $\th$ and no zero mode (nonvanishing
classical solution) for the bosonic partner $y$.
The zero mode solution for $\th$ and the spectrum
of $y$ for $x_9=0$ are explicitly given in \cite{HLW}.
This configuration contains
not only the two longitudinal five-branes
but also stacks of membranes inside the
five-branes. Our arguments will only rely
on the existence of a chiral fermionic
zero mode, so our conclusions do
not depend on the details of the five-brane
configurations.
Since the five-brane
charges for both five-branes are unity,
the Chern character $\frac{1}{4!(2\pi)^4}
\int Tr(F^4)$ is one. \footnote{By duality
this Chern character for the case of a D0-brane
crossing a D8-brane is just the 8-brane charge.}
Hence by the index theorem \cite{AS}, the difference in
the number of fermionic zero modes in the two
chiralities is one.
In fact as there is only one zero mode on a 0-8 string \cite{Pol},
so by duality it must be the case that there is a single
chiral fermionic zero mode
for the two five-branes.

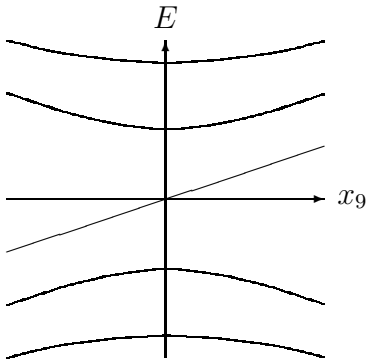
\begin{figure}
\vfill
\begin{center}
\begin{picture}(140,150)(40,40)
\put(40,100){\vector(1,0){120}}
\put(100,40){\vector(0,1){120}}
\put(100,165){\makebox(0,0)[b]{$E$}}
\put(165,100){\makebox(0,0)[l]{$x_9$}}
\put(40,80){\line(3,1){120}}
%\qbezier(40,140)(100,115)(160,140)
\multiput(40,140)(1,-0.333){20}{\line(1,0){1}}
\multiput(60,133.33)(1,-0.25){16}{\line(1,0){1}}
\multiput(76,129.33)(1,-0.167){8}{\line(1,0){1}}
\multiput(84,128)(1,-0.125){12}{\line(1,0){1}}
\put(96,126.5){\line(1,0){8}}
\multiput(104,126.5)(1,0.125){12}{\line(1,0){1}}
\multiput(116,128)(1,0.167){8}{\line(1,0){1}}
\multiput(124,129.33)(1,0.25){16}{\line(1,0){1}}
\multiput(140,133.33)(1,0.333){20}{\line(1,0){1}}
%\qbezier(40,60)(100,85)(160,60)
\multiput(40,60)(1,0.333){20}{\line(1,0){1}}
\multiput(60,66.67)(1,0.25){16}{\line(1,0){1}}
\multiput(76,70.67)(1,0.167){8}{\line(1,0){1}}
\multiput(84,72)(1,0.125){12}{\line(1,0){1}}
\put(96,73.5){\line(1,0){8}}
\multiput(104,73.5)(1,-0.125){12}{\line(1,0){1}}
\multiput(116,72)(1,-0.167){8}{\line(1,0){1}}
\multiput(124,70.67)(1,-0.25){16}{\line(1,0){1}}
\multiput(140,66.67)(1,-0.333){20}{\line(1,0){1}}
%\qbezier(40,160)(100,145)(160,160)
\multiput(40,160)(1,-0.25){16}{\line(1,0){1}}
\multiput(56,156)(1,-0.167){6}{\line(1,0){1}}
\multiput(62,155)(1,-0.125){18}{\line(1,0){1}}
\multiput(80,152.75)(1,-0.0625){20}{\line(1,0){1}}
\multiput(100,151.5)(1,0.0625){20}{\line(1,0){1}}
\multiput(120,152.75)(1,0.125){18}{\line(1,0){1}}
\multiput(138,155)(1,0.167){6}{\line(1,0){1}}
\multiput(144,156)(1,0.25){16}{\line(1,0){1}}
%\qbezier(40,40)(100,55)(160,40)
\multiput(40,40)(1,0.25){16}{\line(1,0){1}}
\multiput(56,44)(1,0.167){6}{\line(1,0){1}}
\multiput(62,45)(1,0.125){18}{\line(1,0){1}}
\multiput(80,47.25)(1,0.0625){20}{\line(1,0){1}}
\multiput(100,48.5)(1,-0.0625){20}{\line(1,0){1}}
\multiput(120,47.25)(1,-0.125){18}{\line(1,0){1}}
\multiput(138,45)(1,-0.167){6}{\line(1,0){1}}
\multiput(144,44)(1,-0.25){16}{\line(1,0){1}}
\end{picture}
\end{center}
\caption{Spectrum of $\th$} \label{spec}
\end{figure}

For a generic configuration of five-branes,
the Hamiltonian and thus the spectrum of
$y$ depend only on $x_9^2$ and not on $x_9$.
But the spectrum of $\th$ is a little bit more
complicated. The Hamiltonian for $\th$ is
\be \label{H}
H_F=2\pi R T_2^M\int \th^{\dag}\H\th,
\eq
where $\H=(iD+\G^{09} x_9)$ with
$D=\sum_{\m=1}^8\G^{0\m}D_{\m}$.
The spectrum of $\H^2$ is
$\{s^2+x_9^2\; |\; s\in\mbox{Spec}(iD)\}$.
This would imply that the spectrum of $\H$
depends only on $x_9^2$ if all $s\neq 0$.
However, if zero is an eigenvalue of $iD$, then
the corresponding eigenvalue(s) of $\H$ can be
$\{x_9\}$, $\{-x_9\}$ or $\{x_9, -x_9\}$.
It was known \cite{HLW} that when $x_9=0$
there is only one zero mode
for $iD$, so the last possibility is ruled out.
Using the chirality of the zero mode
on $T^8$: $\G^1\G^2\cdots\G^8\th=\th$ and
its total chirality: $\G^0\G^1\cdots\G^9\th=\th$,
we find that $\G^{09}\th=\th$.
Thus the correct choice is $\{x_9\}$.
This is the only part of the spectrum that depends
on the sign of $x_9$.
This property makes the fermionic zero mode
behave differently from all other states
in an essential way.
A schematic diagram of
the spectrum of $\th$ is in Fig.\ref{spec}.

When the two five-branes cross each other,
the value of $x_9$ changes its sign.
While all other states are invariant under
the reflection $x_9\rightarrow -x_9$,
the zero mode is not.
This means that the two
five-branes select a prefered direction in
the transverse direction $x_9$. (By duality,
this implies for the D0-D8 case that
the D8-brane is oriented.)
Although one can interchange
the positions of the two diagonal blocks
by a gauge transformation:
\be
X_{\m}\rightarrow UX_{\m}U^{\dag}, \quad
\Psi\rightarrow U\Psi U^{\dag},
\eq
with
\be \label{U}
U=\left(
  \begin{array}{cc}
     0 & \one \\
     \one & 0
  \end{array}
  \right),
\eq
$\th$ is interchanged with
$\thd$ at the same time.
When $\th$ vanishes, the exchange symmetry
of the two five-branes is preserved.
But when the zero mode is present,
exchanging only the five-branes results in a change
in the state of the system.
Also, for the parity transformation $x_9\rightarrow -x_9$
to be a symmetry, it has to be accompanied by
the change $\G^9\rightarrow -\G^9$.

The fermionic zero mode, according to
eq.(\ref{H}), has the energy
\be
E=2\pi R T^M_2 x_9.
\label{E}
\eq
If we compactify the ninth direction to a
circle of radius $R_9$, $x_9$ will be
promoted to $x_9+2\pi n R_9$, where $n$ is
the winding number. Then the energy of the
zero mode is
\be
E_n=2\pi R T^M_2 (x_9+2\pi R_9 n).
\label{nrg2}
\eq
This is equivalent to eq.(3) given in
ref.\cite{BDG}, where it was intepreted
in the string theory context as the
(fermionic) ground state energy for
an open string stretching between
the two branes. Here in M(atrix) theory
we identify it as the zero
mode of fermionic off-diagonal blocks.

\section{Second Quantization and Membrane Creation}
\label{SQ}

The absolute value of the energy (\ref{E})
is the same as that for a membrane stretching
between two five-branes.
It is therefore tempting to relate this
energy to the energy for a created
longitudinal membrane.
For this identification to make sense
it is important to notice that
in M(atrix) theory we are working in an
infinite momentum frame, in which the energy
of a longitudinal degree of freedom
do not have the prefactor of
$R/N$ like the transverse degrees of freedom
\cite{BSS}. Due to translational invariance,
its longitudinal momentum vanishes, so that
$E=M$.

\begin{figure}
\vfill
\begin{center}
\begin{picture}(320,120)(20,40)
\put(40,100){\vector(1,0){120}}
\put(100,60){\vector(0,1){80}}
\put(100,145){\makebox(0,0)[b]{$E$}}
\put(165,100){\makebox(0,0)[l]{$x_9$}}
\put(40,120){\line(3,-1){60}}
\thicklines
\put(100,100){\line(1,0){60}}
\put(100,40){\makebox(0,0)[b]{(a)}}
\thinlines
\put(200,100){\vector(1,0){120}}
\put(260,60){\vector(0,1){80}}
\put(260,145){\makebox(0,0)[b]{$E$}}
\put(325,100){\makebox(0,0)[l]{$x_9$}}
\put(260,100){\line(3,1){60}}
\thicklines
\put(200,100){\line(1,0){60}}
\put(260,40){\makebox(0,0)[b]{(b)}}
\end{picture}
\end{center}
\caption{(a) Energy of the state $|e\ra$, and
(b) energy of the state $|f\ra$ measured relative to the vacuum}
\label{EE}
\end{figure}
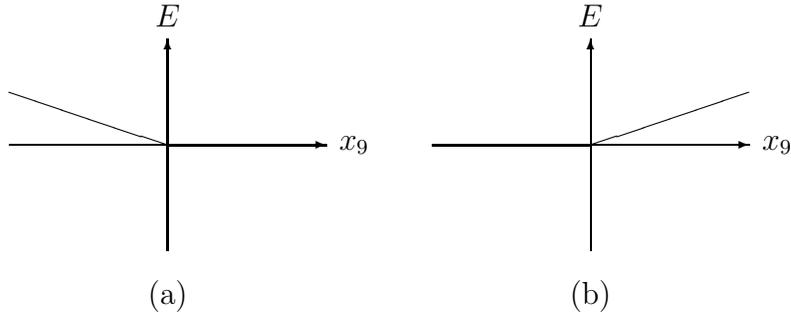

It turns out that quantization is the key to
understand the membrane creation, as well as the half-charges
associated with half-strings in the D0-D8 brane crossing
(see Sec.\ref{FermNum}).
To quantize the fermionic field
$\th$, as a rule we should first fill
all negative-energy states, and define
this (many-body) state as the vacuum of the
fermion system. As for the fermionic zero mode,
it can be either empty or filled. Let us denote
by $|e\ra$ (or $|f\ra$) the state of the fermion
system with all modes below the zero mode filled
(all having negative energies), and with the
zero mode empty (or filled). Which of them is the
vacuum state depends on the sign of $x_9$
since the zero mode energy is proportional to $x_9$.
When $x_9<0$, the state
$|f\ra$ is the vacuum and $|e\ra$ is a hole; when
$x_9>0$, the state $|e\ra$ is the vacuum and
$|f\ra$ is a particle.
The energy of the states measured relative to the vacuum
is given in Fig.\ref{EE}(a) and \ref{EE}(b),
respectively for $|e\ra$ and $|f\ra$.

Let us denote the amplitude of the zero mode by $\chi$,
so that $\th=\chi\th^0$ where $\th^0$ is the zero mode
solution as a function defined on $T^8$ \cite{HLW}.
The canonical quantization of $\chi$ is
realized on the Hilbert space spanned by the two
states $|e\ra$ and $|f\ra$.
Up to constant factors,
$\chi$ and $\chid$ act on the Hilbert space
as annihilation and creation operators, respectively:
$\chi|e\ra=0$, $\chi|f\ra=|e\ra$,
$\chid|e\ra=|f\ra$ and $\chid|f\ra=0$.

Whether $\chi$ is the annihilation operator for a particle
or the creation operator for a hole depends on the sign of $x_9$.
The energies in Fig.\ref{EE} are obtained by using
the usual normal ordering (i.e. putting annihilation operators
to the right of creation operators) for the Hamiltonian
for $x_9<0$ and $x_9>0$ separately.
Energies obtained by normal ordering are those measured
relative to the ground state and thus are always non-negative.

Let us denote the vacuum as $|0\ra$ and
the single-particle excited state
as $|1\ra$ for all $x_9$.
Hence $|0\ra$ is equal to $|e\ra$ for $x_9>0$
but is $|f\ra$ for $x_9<0$.
Similarly, $|1\ra$ is $|f\ra$ for $x_9>0$
and is $|e\ra$ for $x_9<0$,
corresponding to either a particle or a hole.
The difference between the energies of the states $|0\ra$ and $|1\ra$
is the same as the energy of a longitudinal membrane
of length $| x_9 |$.
This suggests that
the states $|0\ra$ and $|1\ra$ represent the situations
with zero and one open membrane stretching between
the five-branes, respectively.

Now we come to the crucial point:
When $x_9$ changes adiabatically,
the zero mode remains either empty or filled
according to quantum adiabatic theorem.
Thus, as $x_9$ crosses
zero from the positive to the negative side,
from Fig.\ref{EE}(a) we see a spontaneous creation
of a hole from the vacuum due to the spectral
flow of the zero mode, while in Fig.\ref{EE}(b) a
spontaneous annihilation of a particle into
the vacuum.
Everytime the energy of the fermionic zero mode
crosses zero a particle or a hole is created or annihilated.
The underlying physics is simply spectral flow
plus filling of the Dirac sea.

Following a similar suggestion \cite{BDG}
in string theory, we interpret what is created
or annihilated, in association with the above
particle (hole) creation or annihilation,
as a longitudinal membrane in the present M(atrix)
theory context. From Fig.\ref{EE}(a) and
\ref{EE}(b) one can see that for the
two five-branes in question, assuming $x_9$
is non-compact, initially there are two
possibilities in the second quantized theory,
either there is none or there is one longitudinal
membrane stretching between them. In the first
(or the second) case, the crossing of the two
five-branes will lead to creation (or annihilation)
of such a membrane.
The necessity of having both possibilities was argued
in ref.\cite{HW} (Fig.9).

\section{Half Fermion Number and Operator Ordering}
\label{FermNum}

In this section, we pay attention to the problem
of the so-called half-strings in the D0-D8 case
\cite{DFK,BGL}. We will see that this problem is
related to the well-known fermion number fractionalization
in the presence of a fermionic zero mode (or in general
mid-gap state) \cite{JR}.

\begin{figure}
\vfill
\begin{center}
\begin{picture}(180,100)(20,40)
\put(40,80){\vector(1,0){120}}
\put(100,40){\vector(0,1){80}}
\put(100,125){\makebox(0,0)[b]{$N_F$}}
\put(165,80){\makebox(0,0)[l]{$x_9$}}
\put(100,100){\line(1,0){60}}
\put(105,105){\makebox(0,0)[lb]{$1/2$}}
\put(165,100){\makebox(0,0)[l]{$|1\ra$}}
\put(40,60){\line(1,0){60}}
\put(35,60){\makebox(0,0)[r]{$|1\ra$}}
\put(105,55){\makebox(0,0)[lt]{$-1/2$}}
\multiput(40,100)(5,0){12}{\line(1,0){3}}
\put(35,100){\makebox(0,0)[r]{$|0\ra$}}
\multiput(100,60)(5,0){12}{\line(1,0){3}}
\put(165,60){\makebox(0,0)[l]{$|0\ra$}}
\end{picture}
\end{center}
\caption{Fermion number operator $N_F$ for the zero mode} \label{NF}
\end{figure}
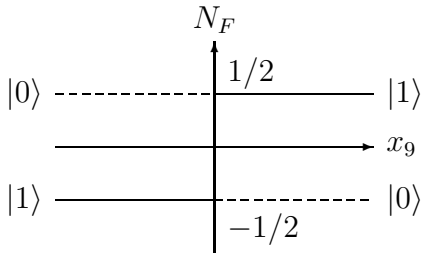

Classically, the conserved fermion number for the zero mode
is $\chid\chi$.
In quantum mechanics, the operators have to be properly ordered
so that it transforms correctly
under the fermion-number conjugation \cite{JR}.
For an ordinary fermionic field the number operator is
\be
(\sum_{\a}b_{\a}^{\dag}b_{\a}-\sum_{\a}d_{\a}^{\dag}d_{\a}),
\eq
where the $b_\a$'s and $d_\a$'s are the annihilation operators
for particles and holes, respectively.
Since the creation of a hole is the same as
the annihilation of a particle in the Dirac sea,
the roles of the $b_{\a}$'s and $d_\a$'s
are symmetric up to a flip of sign
in the fermion number.
At $x_9=0$, the two states $|e\ra$ and $|f\ra$ are degenerate.
To preserve the symmetric roles of $|e\ra$ and $|f\ra$,
\footnote{Note that it is a matter of convention to say
which state is empty or filled,
as well as which operator is the creation
or annihilation operator.}
the fermion numbers for these states should be $\pm 1/2$ \cite{JR}.
It is thus fixed up to a sign to be
\be \label{fermion}
N_F=\frac{1}{2}(\chid\chi-\chi\chid).
\eq
This operator only takes values of $\pm 1/2$ (Fig.\ref{NF}).
It is $-1/2$ for $|e\ra$ and $1/2$ for $|f\ra$.
The value of $N_F$ for the vacuum
changes by one when $x_9$ crosses zero,
showing the need for the creation of a membrane
in order to maintain charge conservation.

Comparing with the dual \IIB system of NS5-brane and D5-brane
\cite{HW}, the charge (\ref{fermion}) is
the total magnetic charge.
The charge of the vacuum $|0\ra$ corresponds to the induced charge
due to one of the five-branes on the other five-brane.
Due to the jump of the induced charge,
conservation of the total magnetic charge
requires the creation of a membrane.

In \cite{DFK} it was shown that in \IIA theory
the induced charge on a D8-brane
by a D0-brane is one half and it jumps to minus one half
when the branes cross.
(See \cite{HW} for the \IIB analogue.)
The charge one half was associated with half a string \cite{DFK}.
By duality we see that this peculiar appearance
of half a string is simply
originated from the ordering of operators in quantum mechanics.
To use the term half a string is in a sense just like saying
that the ground state energy for a simple harmonic oscillator
is due to half a quantum.

As mentioned in Sec.\ref{zero},
interchanging the two five-branes is accompanied by
interchanging $\th$ and $\thd$, which implies that
$N_F\rightarrow -N_F$.
Thus if we say that $N_F$ is the induced charge on the first five-brane,
the induced charge on the second five-brane would be $-N_F$.

In fact, the fermion number for $|0\ra$ jumps by one
when $x_9$ crosses zero for an arbitrary ordering of operators.
When the 9-th direction is compactified,
$x_9$ becomes a $U(1)$ gauge field in a 1+1 dimensional theory,
and $\chi$ becomes a fermionic field
charged with respect to this gauge field.
The jump in the fermion number causes a change in
the number of charged fields
and thus affects the $U(1)$ anomaly
for the 1+1 dimensional field theory
\cite{BDG}.
In an adiabatic process where $x_9$ passes zero,
a membrane is created so that the total charge
is conserved.
The orientation of the membrane is determined by
the charge conservation.

\section{Energy of Zero Mode and Effective Potential}
\label{potential}

Knowing the spectrum of the zero mode (Fig.\ref{spec}),
we can derive the effective potential in the Hamiltonian
approach without much effort. The calculation for the
effective potential is analogous to that for the Casimir effect.
For effective potential between two branes, we are
concerned about the change in the vacuum energy
as $x_9$ varies. For sufficiently small $x_9$, the
dominating contribution to the force between the two
branes comes from the zero mode,
which is constant with respect to $x_9$.
The reason is that except for the zero mode, the spectra of
other modes of $y$ and $\th$ depend only on $x_9^2$,
which always give a force proportional to $x_9$.
The effective potential between the five-branes
is therefore approximately the energy of the zero
mode.

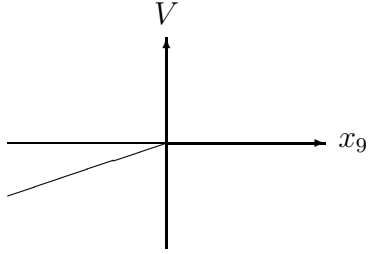
\begin{figure}
\vfill
\begin{center}
\begin{picture}(140,100)(20,60)
\put(40,100){\vector(1,0){120}}
\put(100,60){\vector(0,1){80}}
\put(100,145){\makebox(0,0)[b]{$V$}}
\put(165,100){\makebox(0,0)[l]{$x_9$}}
\put(40,80){\line(3,1){60}}
\thicklines
\put(100,100){\line(1,0){60}}
\end{picture}
\end{center}
\caption{The effective potential between the two five-branes}
\label{V}
\end{figure}

After one integrates over $T^8$
the Hamiltonian operator (\ref{H}) becomes
\be \label{HF}
H_F=2\pi R T_2^M x_9 \chid\chi.
\eq
It follows that the effective potential between the two
five-branes when there is no membrane stretching
between them (corresponding to the state $|0\ra$)
is given by Fig.\ref{V}.
%The energy for the state $|1\ra$ is equal to the sum of
%the potential for $|0\ra$ and the energy due to the
%stretched membrane (Fig.\ref{V1}).
Since the state $|0\ra$ is the ground state for all $x_9$,
it should be the state corresponding to the string theory
calculations in \cite{Lifs,BGL}.

%\begin{figure}
%\vfill
%\begin{center}
%\begin{picture}(140,100)(20,60)
%\put(40,100){\vector(1,0){120}}
%\put(100,60){\vector(0,1){80}}
%\put(100,145){\makebox(0,0)[b]{$V$}}
%\put(165,100){\makebox(0,0)[l]{$x_9$}}
%\put(100,100){\line(3,1){60}}
%\thicklines
%\put(40,100){\line(1,0){60}}
%\put(260,40){\makebox(0,0)[b]{(b)}}
%\end{picture}
%\end{center}
%\caption{The energy of two five-branes with a stretched membrane}
%\label{V1}
%\end{figure}

The operator ordering in (\ref{HF}) agrees with the normal ordering
when $x_9>0$ but is different from
the normal ordering for $x_9<0$.
Normal ordering is used to calculate
the excitation energy relative to the vacuum,
but here we are interested in the variation in the
vacuum energy itself.
One may wonder if one can choose another operator ordering,
for instance, to use
$\frac{1}{2}(\chid\chi-\chi\chid)$
instead of $\chid\chi$ in (\ref{HF}).
This ambiguity of operator ordering
is fixed by requiring the exchange symmetry
mentioned in Sec.\ref{zero}.
When the zero mode is absent, exchanging the five-branes
($Z\leftrightarrow W$) together with a translation in $x_9$
results in the transformation $x_9\rightarrow -x_9$.
As a symmetry this should leave the energy invariant,
so we should have $H_F=0$ for the state $|e\ra$.
Note that the ordering should be independent of $x_9$
in order to exhibit the $x_9$-dependence of the vacuum energy.

Consider the vacuum state $|0\ra$.
When $x_9<0$, the zero mode is filled and
the variation of its energy with respect to $x_9$
gives a repulsive force between the five-branes.
For $x_9>0$, the zero mode is empty
and the force vanishes.
Comparing the energy for $|0\ra$ and the effective
potential between D0 and D8-branes by duality,
we find that indeed $|0\ra$ is the state corresponding to
the string calculations in \cite{Lifs,BGL}.
Note that, contrary to what is suggested in some of the literature,
we believe that the change in force for the state $|0\ra$
when $x_9$ changes sign is not a signal of membrane creation,
rather it is only a result of closed string R-R exchange.
Indeed, the membrane creation is associated with
the jump from $|0\ra$ to $|1\ra$.

Combining Fig.\ref{EE}(a) and Fig.\ref{V},
one easily sees that with the created brane included,
the net potential (and force) between the two five-branes
remains zero when $x_9$ changes from negative to positive,
if initially there is no membrane suspended between them.

String theory calculations \cite{Lifs,BGL}
show that the effective potential between a D0-brane
and a D8-brane is
\be \label{pot}
V=-\frac{1}{2}T_s(1\mp 1)|x_9|,
\eq
in agreement with our results.
The first term above comes from the traces over
the NS and R sectors and the second term
from the R$(-1)^F$ sector of open strings.
The sign depends on whether the D0-brane
is on the left or right of the D8-brane.
Thus the sign flips when the branes cross.
This is identical to what we see in the M(atrix) theory.
When $x_9$ crosses zero for the state $|0\ra$, only the zero mode
will change its fermion number by one (see Fig.\ref{NF})
and cause a change in sign
of the R$(-1)^F$ sector.
The string theory calculation also shows that
one has to attribute part of the effective potential
to gravitons and dilatons so that the rest is due to
the contribution of half a string.
In M(atrix) thoery we no longer distinguish
contributions from the NS, R or R$(-1)^F$ sectors,
\footnote{However, it appears that the two terms in (\ref{pot})
correspond to the decomposition of $\chid\chi$ in (\ref{HF})
as $\frac{1}{2}(\chid\chi+\chi\chid)+\frac{1}{2}(\chid\chi-\chi\chid)$
for the state $|0\ra$.}
but we can understand interactions solely
in terms of off-diagonal blocks.
While half a string can hardly be physical in string theory,
we now have a better understanding of
the total effect of interactions in M(atrix) theory.

The low energy effective potential between the
five-branes can also be calculated by integrating out
$y$ and $\th$. Since the action for $y$ and $\th$
is the same as the one for 0-8 strings (in the
strong coupling limit), we can apply the
results of \cite{Pier}. Though it was claimed
in \cite{Pier} that the M(atrix) theory result
does not agree with the string theory calculation
\cite{Pier,Lifs}, we note that this discrepancy
is due to an assumption
that the potential is independent of the sign
of $x_9$. Without this assumption the calculation
in \cite{Pier} would have given a result
consistent with our calculation of the vacuum energy.

\section{Remarks} \label{remark}

We conclude this paper by a few remarks.

\begin{enumerate}

\item We find it much easier to calculate
the effective potential and to understand
the creation of an open membrane
in M(atrix) theory than in string theory.
What is essential to our arguments in
the M(atrix) theory is
the generic feature of the spectrum (Fig.\ref{spec})
for the off-diagonal blocks.
The quantization of the fermionic zero mode associates
the creation of the open membrane
in its ground state with the creation and annihilation
operators $\chid, \chi$.
The notion of ``half a membrane'' or ``half a string''
is understood as a result of operator ordering
appropriate in the presence of the fermionic zero mode.

\item We used the zero mode in the off-diagonal blocks
to describe the creation of an open membrane,
while the five-branes are given by the diagonal blocks.
This is to be contrasted with other descriptions
of open membranes by using $SO(N)$ matrices \cite{SO}
or by modifying the closed membrane configuration \cite{Li}.
%Although it is sufficient for the purpose of this paper,
%we still need a description of open membranes
%using diagonal blocks in order to discuss, for instance,
%interactions between membranes stretching between parallel five-branes.

\item The intriguing behavior of the zero mode
is due to the fact that the zero mode is chiral
in the transverse direction.
One may apply similar arguments to the generic case
of a D$p$-brane  and a D$p'$-brane.
But except those dual to the five-branes discussed above,
there is no similar phenomenon because the fermionic zero modes
of both chiralities in the transverse direction
are paired.

\item Applying our discussions to the case of one D0-brane
in the presence of two D8-branes
considered in \cite{BGL}, we find for the ground state
that the forces in the
three regions divided by the two D8-branes are from
left to right $-2T_s$, $-T_s$ and $0$, respectively.
It is simply the superposition of the forces
on the D0-brane due to the two D8-branes.
However in \cite{BGL} it was argued that the forces
from left to right should be $-2T_s$, $0$ and $0$, respectively.
In their arguments they used the cancellation
between two half-strings associated with the two D8-branes
as the total induced charge vanishes
when the D0-brane lies between the two D8-branes.
However in M(atrix) theory the five-branes are naturally
associated with Chan-Paton factors, thus the two
half-strings with different Chan-Paton factors can not cancel.

\end{enumerate}

\section{Acknowledgment}

We thank Igor Klebanov for helpful comments. Y.S.W.
thanks Japan Society for the Promotion of Sciences for an
Invitational Fellowship and Institute for Solid State Physics,
University of Tokyo for warm hospitality, where part of work
was done.
This work is supported in part by
U.S. NSF grant PHY-9601277.

\end{document}